\documentclass[12pt,preprint]{aastex}

\title{Evolution of the Cluster Mass and Correlation Functions in 
a $\Lambda$CDM Cosmology}

\author{Joshua D. Younger\altaffilmark{1}, Neta A. Bahcall\altaffilmark{1}, and Paul 
Bode\altaffilmark{1}}
\altaffiltext{1}{Princeton University Observatory, Princeton NJ 08544}

\begin{document}

\begin{abstract}
The evolution of the cluster mass function and the cluster 
correlation function
from $z=0$ to $z\approx 3$ are determined using $\sim$$10^6$ clusters 
obtained from high-resolution simulations of the current best-fit 
$\Lambda$CDM cosmology ($\Omega_m=0.27$, $\sigma_8=0.84$, $h$ $=0.7$). The 
results provide predictions for comparisons with future observations of high 
redshift clusters. A comparison of the predicted mass function of low redshift
clusters with observations from early Sloan Digital Sky Survey data, 
and the predicted abundance of massive distant clusters with observational 
results, favor a slightly larger amplitude of mass fluctuations ($\sigma_8 
\sim 0.9$) and lower density parameter ($\Omega_m \sim 0.2$); these values 
are consistent within 1-$\sigma$ with the current observational and model 
uncertainties. The cluster correlation function strength
increases with redshift for 
a given mass limit;
the clusters were more strongly correlated in the 
past, due to 
their increasing bias with redshift---
the bias reaches 
$b\sim 100$ at $z$=2 for $M>5\times 10^{13} h^{-1}M_{\odot}$ clusters.
The richness-dependent cluster 
correlation function, represented by the correlation scale versus cluster 
mean separation relation, $R_0 - d$, is generally consistent with 
observations. 
This relation can be approximated as $R_0=1.7d^{0.6} h^{-1}$ Mpc 
for $d \sim 20 - 60$ $h^{-1}$ Mpc. The $R_0 - d$ relation 
exhibits surprisingly 
little evolution with redshift for $z < 2$;
this can provide a new test of the current 
$\Lambda$CDM model when compared with future observations of high redshift 
clusters.
\end{abstract}

\section{Introduction}

Clusters of galaxies, the most massive virialized structures in the universe, 
provide vital information about large-scale structure of the universe and 
place powerful constraints on cosmology (Bahcall 1988; Peebles 1993;
Carlberg et al. 1997; Rosati, Borgani \& Norman 2002;
Henry 2004; and references therein).  
The abundance of clusters 
as a function of mass (i.e. the cluster mass function) and 
the evolution of this abundance
with redshift are sensitive probes of both the present day density 
parameter ($\Omega_m$) and the amplitude of mass fluctuations ($\sigma_8$); 
this provides a powerful test of the cosmological model (Peebles, Daly, \& 
Juszkiewicz 1989; Henry \& Arnaud 1991; Bahcall \& Cen 1992; Oukbir \& 
Blanchard 1992; Barlett \& Silk 1993; White \textit{et al.} 1993; Viana \& 
Liddle 1996; Eke, Cole, \& Frenk 1996; Pen 1996; Cen 1998;
Ikebe \textit{et al.} 2002; Seljak 2002; Bahcall \textit{et al.} 2003b;
and references therein).

The spatial distribution of clusters of galaxies serves as a complementary 
test of the cosmological model;  this cluster distribution is often described 
by the two-point cluster correlation function.  The amplitude of the 
correlation function offers a strong test of the cosmology (Bardeen, Bond, \& 
Efstathiou 1987; Bahcall \& Cen 1992; Mann, Heavens \& Peacock 1993; Holtzman 
\& Primack 1993; Croft \& Efstathiou 1994; Borgani \textit{et al.} 1995).  In 
fact, it was the unexpectedly strong observed cluster correlations (Bahcall 
\& Soneira 1983) that provided the first evidence against the then standard 
$\Omega_m = 1$ SCDM models (Bahcall \& Cen 1992; Croft \textit{et al.} 1997; 
Borgani, Plionis, \& Kolokotronis 1999; Governato \textit{et al.} 1999; 
Colberg \textit{et al.} 2000; and references therein).  In addition, the 
evolution of the cluster correlation function with redshift, which has 
received comparatively less attention in the literature, is sensitive to the 
cosmology.

Taken together, the cluster mass and correlation functions provide two of 
the most powerful constraints on cosmological models.  Predictions have
become increasingly robust with larger and higher resolution cosmological 
simulations, made possible by recent growth in computing power combined with 
more sophisticated algorithms.  
Considerable progress has also been made
on the observational front to
determine cosmological parameters.  Combining the recent WMAP
data with finer-scale CMB experiments plus galaxy and Ly-$\alpha$
forest data, Spergel \textit{et al.} (2003) determined a
best-fit power law $\Lambda$CDM model.  While further data will
refine this model, the differences are likely to be small.
As a further check of this model,
we present in this work the 
simulated mass and correlation functions of clusters of galaxies, and their 
evolution with redshift, determined from mock sky survey cluster catalogs 
generated from a $\Lambda$CDM simulation of the current best-fit cosmological 
model (Spergel \textit{et al.} 2003).  We compare our results to the most 
recent observations, and lay the groundwork for comparison with future 
cluster observations at both low and high redshift.  Such comparisons will 
provide important tests of the current cosmology, and will enable further 
improvements in the determination of cosmological parameters.
%The end result is expected to be a complete, self-consistent theory of the 
%history of structure formation in our universe from recombination to the 
%present day, testing the predictive power of the current cosmological model.

\section{The Cluster Mass-Function and Its Evolution}

For the simulation parameters we took those determined by
Spergel \textit{et al.} (2003) from the new WMAP data on
the largest scales, supplemented by other CMB experiments, galaxy
surveys, and Ly-$\alpha$ forest observations on smaller scales.
Assuming a spatially
flat power law $\Lambda$CDM universe, these are:
matter density $\Omega_m =0.27$, 
cosmological constant $\Lambda =0.73$,
power spectrum amplitude $\sigma_8 =0.84$ 
, spectral index $n_s=0.96$, 
and $h=0.7$ where $H_0=100h$ km-s$^{-1}$-Mpc$^{-1}$.
The simulation used the TPM code (Bode \& Ostriker 2003) to evolve
$N=1260^3\approx 2\times 10^9$ particles in a periodic box 1500$h^{-1}$ Mpc on 
a side. 
The particle 
mass is $m_p = 1.26 \times 10^{11} h^{-1} M_\odot$, and a cubic spline 
softening length of $17$ $h^{-1}$kpc was introduced. 
The simulation is discussed in more detail in 
Hopkins, Bahcall \& Bode (2005).

Particle positions in a light cone covering one octant of
the sky out to redshift $z=3$ were saved to disk; snapshots
of the entire simulation volume were also saved.
Dark matter halos, which would house clusters of galaxies, were 
identified from the particle position data using a Friends-of-Friends (FOF) 
percolation algorithm, with a linking length of $b = 0.2$ times the mean 
particle separation (Lacey \& Cole 1994).  The 
cluster center was defined as the location of the most bound particle.  
Clusters identified using linking length parameters of $b = 0.16$ and $0.25$ 
were examined for comparison, yielding similar results.

The mass function (MF) of clusters, $n(>$$ M)$, represents the number density 
of clusters with mass greater than $M$.  
The constraints which the present day cluster MF places 
on the mean density 
parameter of the universe ($\Omega_m$) and the amplitude of mass fluctuations 
($\sigma_8$) are partially
degenerate in $\Omega_m-\sigma_8$.  
Observations of the present day cluster MF
have established that
$\sigma_8 \Omega_m^{0.5} \approx 0.5$ (Henry \& Arnaud 1991; Bahcall \& Cen 
1992; White, Efstathiou, \& Frenk 1993; Eke, Cole, \& Frenk 1996; Viana \& 
Liddle 1996; Kitayama \& Suto 1997; Pen 1998).  This
degeneracy can be broken 
by studying the evolution of the cluster MF with redshift (Peebles, Daly, \& 
Juszkiewicz 1989; Oukbir \& Blanchard 1992, 1997; Eke, Cole, \& Frenk 1996; 
Viana \& Liddle 1996; Bahcall, Fan, \& Cen 1997; Carlberg, Morris, Yee, \& 
Ellingson 1997; Henry 1997, 2000; Bahcall \& Fan 1998; Eke \textit{et al.} 
1998; Donahue \& Voit 1999).   Cluster evolution is exponentially dependent 
on $\sigma_8^2$ (Bahcall, Fan, \& Cen 1997; Bahcall \& Bode 2003), and as a 
result, the most massive clusters evolve strongly in a low-$\sigma_8$, 
$\Omega_m = 1$ universe, producing a very low abundance of massive clusters 
at $z > 0.5$.  Conversely, the evolution of rich clusters is significantly 
weaker in a $\sigma_8 \approx 1$ low-$\Omega_m$ universe, with a considerably 
higher cluster abundance at $z > 0.5$ 
as compared to lower-$\sigma_8$ models.

The simulated cluster MF was determined from the light cone outputs
using cluster masses 
calculated according to typical masses used by observers,
including: mass within 
fixed radii (relative to the center of the cluster)
of $0.5$ $h^{-1}$Mpc comoving ($M_{0.5}$), 1.5 $h^{-1}$Mpc 
comoving ($M_{1.5}$), 0.6 $h^{-1}$Mpc physical ($M_{0.6}$),
and also mass within a radius containing a mean overdensity of 200 relative 
to the critical density ($M_{200}$).  Minimum mass cutoffs were 
chosen in order to ensure the 
completeness of the cluster sample: 
$M_{0.5} > 1.6 \times 10^{13} h^{-1} M_{\odot}$; 
$M_{1.5} >5  \times 10^{13} h^{-1} M_{\odot}$; 
$M_{0.6} > 3 \times 10^{13} h^{-1} M_{\odot}$; and
$M_{vir} > 1.75 \times 10^{13} h^{-1} M_{\odot}$. 
The evolution of the 
cluster MF for $M_{0.5}$ is presented in Figure 1.
These results can be used for comparison of predictions
of the current cosmological model with  future observations
of high redshift clusters.

The cluster MF for $M_{0.6}$ and $z = 0.1-0.2$ was compared to the observed 
early Sloan Digital Sky Survey cluster MF for the same redshift and mass 
range (Bahcall \textit{et al} 2003b; see also Bahcall \textit{et al.} 2003a).
The results are presented in Figure 2;  also shown are the best analytic 
model fits.  
The observed data follow the same shape as the LCDM MF,
but are systematically offset to slightly lower masses. This suggests
that either a
bias of $\sim$20\% ($\sim$1-sigma) exists in the observed 
cluster mass calibration (the mass
calibration is expected
to improve for the larger upcoming sample of SDSS clusters) or that
$\Omega_m$ is
somewhat lower than used in the simulation. The best-fit parameters to
the observed
SDSS MF are $\Omega_m\approx 0.2$ and $\sigma_8\approx 0.9$, 
as shown in Figure 2.
These best-fit parameters are consistent with the current model
parameters within
the combined observational and model uncertainties.

The predicted abundance evolution of high mass 
clusters can be compared
to observations at higher redshift.  We use the mass threshold
$M_{1.5} > 8 \times 10^{14} h^{-1} M_{\odot}$ and observed abundances 
of Bahcall \& Bode (2003);  the results are presented in Figure 3.  At high 
redshift, the model predicts considerably fewer clusters than observed.  This 
suggests that the amplitude of mass fluctuations $\sigma_8$ 
is larger than 0.84.  
Bahcall \& Bode (2003) found a best fit value of $\sigma_8 = 0.9\pm 0.1$ for 
these data.  These results are consistent with the current cosmological 
parameter ($\sigma_8 = 0.84\pm0.04$) within the combined observational and 
model uncertainties.
Both the local cluster MF and the evolution of massive clusters at high
redshift thus suggest $\sigma_8\approx 0.9$ and $\Omega_m \approx 0.2$. 
These results indicate a $\sigma_8$
value at the high end of the current best-fit model allowed range,  with
$\Omega_m$ at
the lower end of the accepted range;  both values are consistent within
1-sigma with the recent CMB and large-scale structure observations.  Further
observations are needed to
narrow down the precise value of these cosmological parameters.

\section{The Cluster Correlation Function and Its Evolution}

The cluster correlation function (CF) is a statistical measure of how 
strongly clusters of galaxies cluster as a function of scale.  The 
probability of finding a pair of clusters in volumes $V_1$ and $V_2$, as a 
function of pair separation ($r$) is
\begin{equation}
dP = n^2 (1 + \xi_{cc}(r)) dV_1 dV_2,
\end{equation}
where $n$ is the mean number density of clusters, and $\xi_{cc}(r)$ is the 
cluster CF.   The spatial distribution of clusters of galaxies described 
by the cluster CF is sensitive 
to cosmological parameters
(e.g. Bahcall \& Cen 1992;
Borgani, Plionis, \& Kolokotronis 1999; Colberg \textit{et al.} 2000;
and references therein).

Observationally, the cluster CF is an order of magnitude stronger than that 
of individual galaxies: typical galaxy correlation scales are 
$\sim 5 h^{-1}$Mpc, 
as compared to $\sim$$20 - 25 h^{-1}$Mpc for the richest 
clusters (Bahcall \& Soneira 1983; Klypin \& Kopylov 1983; see also Bahcall 
1988; Huchra \textit{et al.} 1990; Postman, Huchra, \& Guller 1992; Bahcall 
\& West 1992; Peacock \& West 1992; Dalton \textit{et al.} 1994; Croft 
\textit{et al.} 1997; Abadi, Lambas, \& Muriel 1998; Lee \& Park 1999; 
Borgani, Plionis, \& Kolokotronis 1999; Collins \textit{et al.} 2000; 
Gonzalez, Zaritsky, \& Wechsler 2002; and references therein).  Furthermore, 
the strength of the CF increases with cluster richness and mass (Bahcall \& 
Soneira 1983).  As a result, the rarest, most massive clusters exhibit the 
strongest correlations.  

The richness dependence of the cluster CF has been confirmed observationally 
(see references above), and explained theoretically (Kaiser 1984; Bahcall \& 
Cen 1992; Mo \& White 1996; Governato \textit{et al.} 1999; Colberg 
\textit{et al.} 2000; Moscardini \textit{et al.} 2000; Sheth, Mo, \& Tormen 
2001; and references therein).  However, these analyses have been done at low 
redshift, $z < 0.5$.  With observational data becoming available at higher 
redshifts, the expected evolution of the cluster correlation function is 
increasingly important as an
independent test of the cosmological model.  Analytic 
approximations to the evolution of cluster halo abundance, bias, and 
clustering have yielded some promising results (Mann, Heavens, \& Peacock 
1993; Mo \& White 1996, 2002; Sheth, Mo, \& Tormen 2001; Moscardini 
\textit{et al.} 2001).  However, numerical simulations 
can provide the most 
reliable comparison between theory and observations.  We determine the 
expected cluster CF and its evolution for the current best-fit $\Lambda$CDM 
model using light-cone outputs from the N-body cosmological simulation 
discussed in \S 2.  When compared with recent observational results at low 
redshift, this provides a test of the current cosmological model; at the same 
time the evolution of the $\Lambda$CDM CF provides detailed predictions for 
comparison with future observations of high redshift clusters.

The cluster CF was calculated as a function of separation using
$\xi_{cc}(r) = F_{DD}(r)/F_{RR}(r) - 1$,
where $F_{DD}(r)$ is the frequency of cluster pairs with comoving separation 
$r$, and $F_{RR}(r)$ is the frequency of pairs in a random catalog. The 
cluster CF was calculated for different mass thresholds for $M_{0.5}$ and 
$M_{200}$ in several redshift bins: $z$= 0--0.2, 0.45--0.55, 
0.9--1.1, 1.4--1.6, and 1.9--2.2.  
Because of the rapidly decreasing abundance of the 
most massive clusters with redshift, their CF is studied only at lower 
redshifts.  Examples of the cluster CF at different redshifts are presented 
in Figure 4.

The cluster CF for each redshift bin and mass threshold was fit to a power 
law of the form
$\xi_{cc}(r) = (r/R_{0})^{-\gamma}$,
where $\gamma$ is the correlation slope and $R_0$ is the correlation scale.  
The fits were done over the linear range of the cluster CF 
($r \leq 50 h^{-1}$Mpc) 
for both a fixed slope of $\gamma =2$, and 
for $\gamma$ as a free 
fitting parameter.  The results are similar for both methods, with the best 
fit free slope $\gamma \approx 1.8$ for $z < 1.5$.  The slope is mildly 
richness-dependent, with more massive clusters showing a slightly steeper 
slope.  The evolution of $R_0$ and $\gamma$ with redshift is presented in 
Figure 5.

The correlation scale $R_0$ increases both with cluster mass 
and with redshift 
(see Figure 5).  The steepening slope at high redshifts causes $R_0(z)$ to be 
slightly lower for a free $\gamma$ as compared to that of a fixed $\gamma = 
2$, but the evolutionary trend remains the same.  The evolutionary increase 
of the cluster correlation scale with redshift is stronger for more massive 
clusters at higher redshift.  For example, clusters with $M_{0.5} > 1.6 
\times 10^{13} h^{-1}M_{\odot}$ have $R_0 \approx 10 h^{-1}$Mpc at 
$z=0.15$, $R_0 \approx 13 h^{-1}$Mpc at $z=1$, and 
$R_0 \approx 17 h^{-1}$Mpc at $z=2$; 
while clusters with 
$M_{0.5} > 3.0 \times 10^{13} h^{-1}M_\odot$ 
have $R_0 \approx 12 h^{-1}$Mpc at $z = 0.15$, $R_0 
\approx 17 h^{-1}$Mpc at $z=1$, and $R_0 \approx 29 h^{-1}$Mpc at 
$z=2$.  The free slope fits yield similar results.

The cluster CF (using $M_{FOF}$)
and the CF of the dark matter particles were also determined for 
simulation box snapshots at various redshifts, 
and fit to a power law with a fixed 
slope $\gamma = 2$.  The correlation 
scale of clusters ($R_0^{cl}$), as before, increases with redshift.  By 
contrast, the correlation scale of the mass ($R_0^{m}$) decreases with 
redshift, as expected.  This is due to the fact that clusters of a 
given comoving mass represent higher density peaks of the mass distribution 
as the redshift increases, and thus exhibit enhanced clustering with 
increasing redshift (see also Cole \& Kaiser 1989; Mo \& White 1996, 2002; 
Sheth, Mo, \& Tormen 2001; Moscardini \textit{et al.} 2001).  Figure 6 shows 
the evolution of the ratio $R_0^{cl}/R_0^{m}$ with redshift, which 
follows the evolution of bias from $z =$ 0 to 2, where bias is defined 
as $b^2 = \xi_{cc}/\xi_{mm}\approx (R_0^{cl}/R_0^{m})^\gamma$.  As seen in 
Figure 6, the ratio $R_0^{cl}/R_0^m$, and thus the bias $b$, increases from 
$\sim$3 to $\sim$100 (for $M_{FOF}>5\times 10^{13} h^{-1}M_{\odot}$ 
clusters) as the redshift increases from $z=0$ to 2.  Other mass clusters 
show a similar trend.

Another useful approach to studying the evolution of the cluster 
correlation function is the $R_0-d$ relation, where $R_0$ is the fitted 
correlation scale and $d$ is the mean intercluster comoving separation. 
For a larger mass limit objects are less abundant,  and thus their mean
separation $d$ increases.   These objects are also more
biased, so 
an increasing $R_0$ with $d$ is observed (Bahcall 
\& Soneira 1983; Szalay \& Schramm 1985; Bahcall 1988; Croft \textit{et al.} 
1997; Governato \textit{et al.} 1999; Bahcall \textit{et al.} 2003c; 
Padilla \textit{et al.} 2004; and 
references therein).  Thus, the evolution of the $R_0-d$ relation allows us 
to investigate the change in correlation strength with cluster mass and 
with
redshift.  We present the $R_0-d$ relation (using $M_{0.5}$ mass thresholds 
and a fixed $\gamma$) for several redshift bins 
in Figure 7. The results for a free 
$\gamma$ are shown in Figure 8.

The resulting $R_0-d$ relation shows a surprising behavior;  there is 
essentially no evolution with redshifts 
for $z<2$ (using $M_{0.5}$ thresholds).  
Mass thresholds measured within a fixed overdensity ($M_{200}$ and $M_{FOF}$) 
show slightly more evolution, but still surprisingly little.  This redshift 
invariant $R_0-d$ relation provides a powerful test of the cosmology when 
compared with upcoming observations of high redshift clusters.

Why is the $R_0-d$ relation invariant with redshift?
The invariance appears to be partly due to the  relative
constancy of the cluster mass hierarchy with redshift.
That is, the halos which are the most massive at
an early time tend to remain part of the
population of the most massive objects.
Consider a given comoving volume of space.
The majority, but not all, of the $N$ most massive clusters at $z=0$ are
among the
$N$ most massive clusters at higher redshift. The clusters that are not in
the top $N$ at
higher redshift are, in turn, clusters lower down on the mass scale.  In
such a  case,
if the sample of clusters at a given mean comoving separation $d$ is not
dramatically
different at different redshifts, it will yield similar correlation
strengths $R_0$,
assuming that the clusters have not moved significantly over that time
period.
To confirm this, we select clusters using $M_{0.5}$ from box snapshots 
at $z$=0, 0.94, 1.4, and 2.0.  
A cluster at high redshift is considered to be 
the `matched' system if it is within 3 $h^{-1}$Mpc physical 
separation of its 
position at $z=0$.  Using a fixed comoving mean separation
$d = (N/V)^{-1/3}$ at all redshifts, the 
$N_m$ clusters that match with their $z=0$ counterpart are kept, and the 
remaining $N-N_{m}$ clusters are selected randomly from the next $N$ clusters 
down the mass ladder.  Using this new distribution, we find that the 
correlation scale $R_0(d,z)$ is indeed
nearly constant with $z$ for a given $d$ 
(for 
$d\lesssim 90 h^{-1}$Mpc and $z<2$).  
When looking along the past light cone,
the $R_0-d$ relation is thus nearly independent of 
redshift for $z<2$
because the majority of clusters at different redshifts 
represent a similar population 
along the same filamentary hierarchical structure,
near the top of the cluster mass hierarchy.

We compare the predictions of the $\Lambda$CDM model with current $R_0-d$ 
observations in Figure 9, using $M_{0.5}$ 
thresholds and a fixed correlation slope of $\gamma =2$ 
(for details on the observations see Table 1 of Bahcall 
\textit{et al.} 2003c).  All the observed correlation scales 
($R_0$) and mean separations ($d$) were converted to comoving scales in a 
$\Lambda$CDM cosmology.  The band in Figure 9 represents the simulated 
$R_0-d$ relation (with 1-$\sigma$ range) at $z = 0- 0.3$ (to match the 
redshift range of the observations).  An analytic approximation to the 
$\Lambda$CDM $R_0-d$ relation for 
$20 \leq d \leq 60 h^{-1}$Mpc,
presented by the dashed curve, is
\begin{equation}
R_0 = 1.7 \left( \frac{d}{h^{-1}\mathrm{Mpc}} \right)^{0.6} \; 
	   h^{-1} \: \mathrm{Mpc}
%R_0 = 1.7\times d_{\mathrm{Mpc}}^{0.6} \; \textrm{h}^{-1} \: \mathrm{Mpc}
%%R_0 = 1.74 \times \left (\frac{d}{1 \: \mathrm{h}^{-1} \: \mathrm{Mpc}} 
%%\right )^{0.59} \; \textrm{h}^{-1} \: \mathrm{Mpc}
\hspace{1cm} .
\end{equation}
The results show a general agreement between the $\Lambda$CDM model and 
observations.  The optically selected cluster samples agree with the model 
within 1-$\sigma$. There is, however, a wide scatter at the high-$d$ end, 
especially when X-ray selected clusters are included;  the X-ray selected 
clusters seem to suggest a somewhat larger $R_0$ than optically selected 
clusters.  Due to this large scatter, the correlation function cannot yet 
provide a high-precision determination of cosmological parameters.  It does, 
however, clearly rule out high-$\Omega_m$ models, as their far weaker 
correlations are inconsistent with the data.  The strong correlations 
suggested by the current X-ray clusters will be tested with future 
observations of X-ray selected cluster samples; this should clarify whether 
or not the X-ray clusters are consistent with the optical data and with the 
best-fit $\Lambda$CDM cosmology.

\section{Conclusions}

We use light cone outputs from a large-scale simulation of the currently 
best-fit $\Lambda$CDM cosmological model to generate mock sky survey cluster 
catalogs that can be readily compared with observations.  The catalogs 
were used
to determine the present day cluster mass and correlation functions, which 
together constitute a sensitive test of the cosmological model.  We determine 
the evolution of the cluster mass function from $z=0$ to 
$z=3$, and the 
evolution of the cluster correlation function from $z=0$ to $z=2$.  These 
results provide predictions of the current cosmological model for comparison 
with future observations of high redshift clusters.  Such comparisons can be 
used to test the current model and provide new 
and independent constraints on both the 
cosmological density parameter, $\Omega_m$, and the amplitude of mass 
fluctuations, $\sigma_8$.

The simulated cluster mass function at low redshift ($z$=0.1--0.2) is 
compared with
the early Sloan Digital Sky Survey cluster mass function (SDSS: 
Bahcall \textit{et al.} 2003b).  The $\Lambda$CDM mass function predicts 
somewhat higher abundances than are observed.   This suggests that either a 
small bias ($\sim$20\%) exists in the observed cluster mass calibration, or a 
somewhat
lower value of the cosmological density parameter is needed 
($\Omega_m \approx 0.2$).   The results, however, are consistent with the 
current
cosmology within the combined observational uncertainties.
The evolution of the most massive clusters is 
exponentially dependent on $\sigma_8^2$ (Bahcall \& Fan 1998), and therefore 
can be used to break the $\Omega_m-\sigma_8$ degeneracy that exists in the 
low-redshift mass function.  We find that the $\Lambda$CDM model predicts a 
considerably lower abundance of distant ($z > 0.5$) massive clusters than 
observed.  This suggests a
normalization of $\sigma_8\approx 0.9$.

We determine the cluster correlation function in the $\Lambda$CDM cosmology 
as a function of cluster mass threshold, summarized in the $R_0-d$ relation 
(Figure 7--8).  We find good agreement with observations of optically selected 
clusters (Figure 9); 
X-ray selected clusters appear to suggest somewhat increased 
cluster correlations.  The wide scatter in the observational data at 
high-$d$, makes it difficult at this time to provide precise constraints on 
the cosmology.  The data are, however, in good agreement with the current 
$\Lambda$CDM model.

We determine the evolution of the richness dependent cluster 
correlation function in the $\Lambda$CDM cosmology (Figures $4-8$).
For a given mass limit, the correlation strength increases
with redshift.
Surprisingly, the 
$R_0-d$ relation shows no significant evolution out to $z < 2$.  This 
surprising result provides a new, independent test of the current 
cosmological model when compared with future observations of high redshift 
clusters.

\acknowledgments
We thank J.P. Ostriker, D. Spergel, and M. Strauss for
helpful discussions.
The computations were performed on the National Science Foundation
Terascale Computing System at the Pittsburgh Supercomputing Center. 
This work was supported in part by NSF grant AST-0407305.

\begin{figure}
\label{fig1}
\plotone{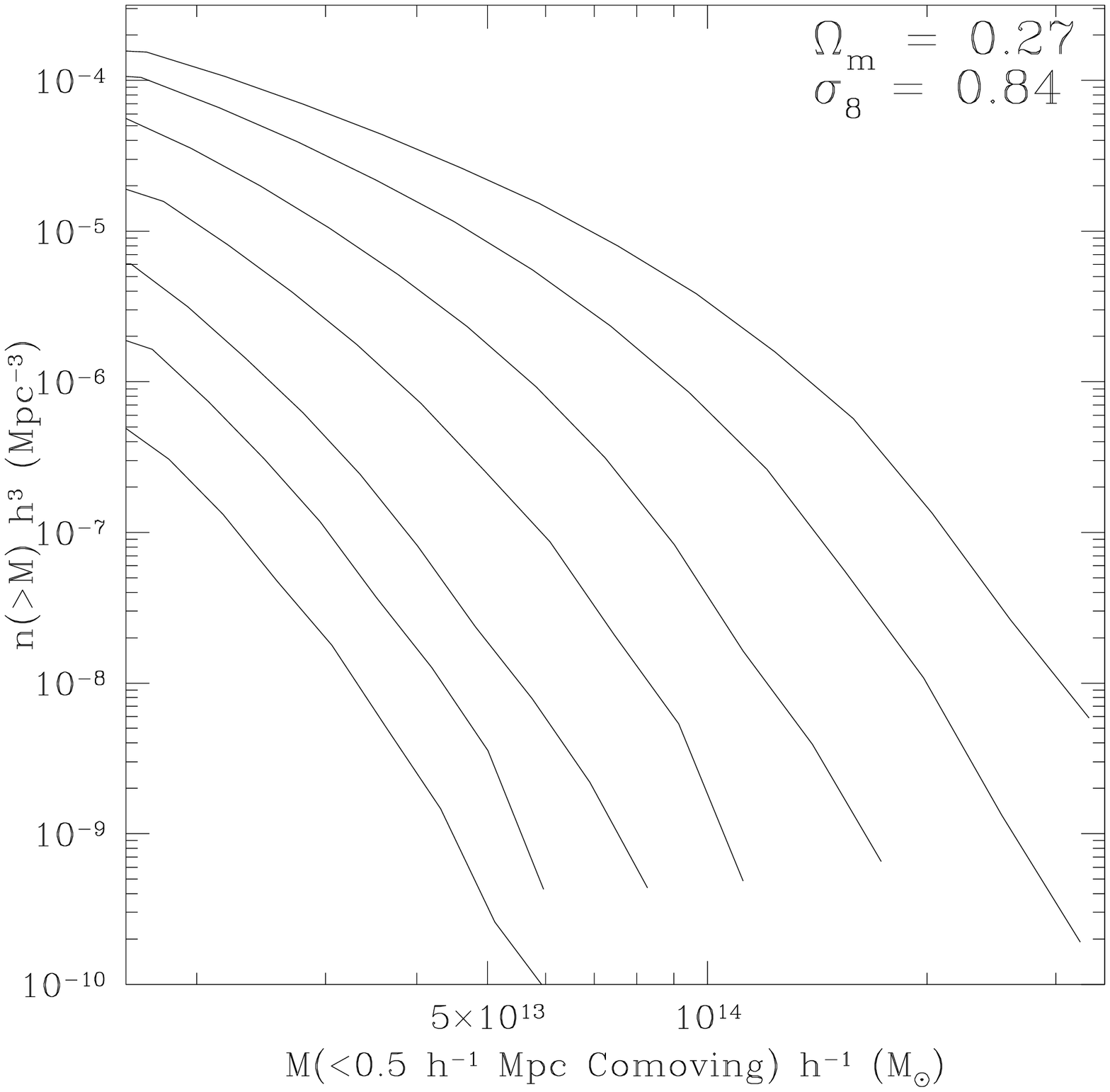}
\caption{The simulated cluster mass function for $\Lambda$CDM, in redshift 
bins $z$= 0.0--0.15; 0.45--0.55; 0.95--1.05; 1.45--1.55; 1.9--2.1; 
2.2--2.5; and 2.5--3.0 (top to bottom).  The mass $M_{0.5}$
is measured within a fixed 
comoving radius 0.5 $h^{-1}$Mpc from the cluster center.}
\end{figure}

\begin{figure}
\label{fig2}
\plotone{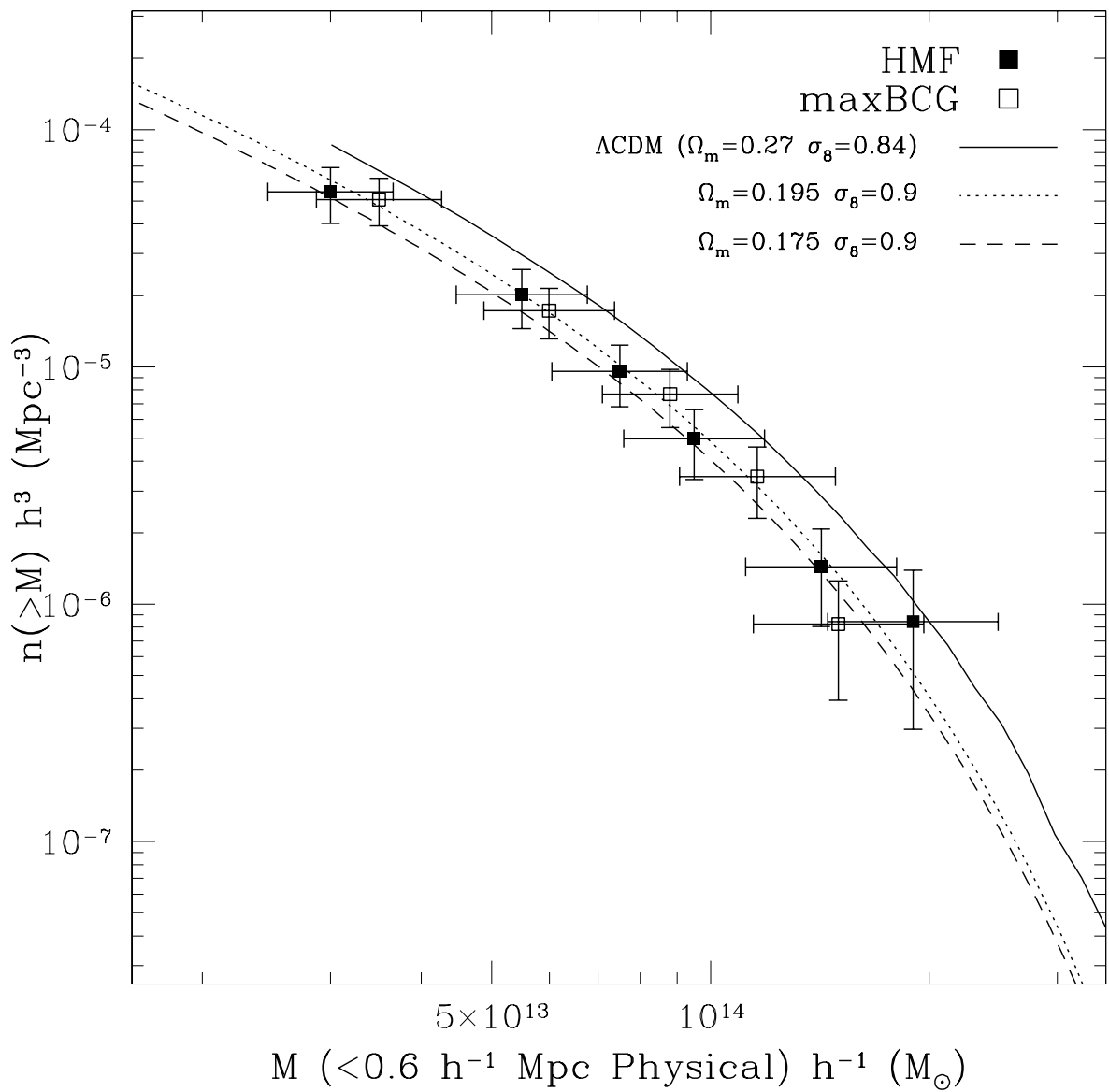}
\caption{A comparison of the  SDSS observed cluster MF and analytic best fits 
for the HMF (filled square, dashed line) and maxBCG (open square, dotted 
line) selection techniques (Bahcall \textit{et al.} 2003a,b) and the current 
$\Lambda$CDM simulation predictions for $z = 0.1 - 0.2$.  All masses are 
defined within a physical radius of 0.6 $h^{-1}$ Mpc of the cluster center.
}
\end{figure}

\begin{figure}
\label{fig3}
\plotone{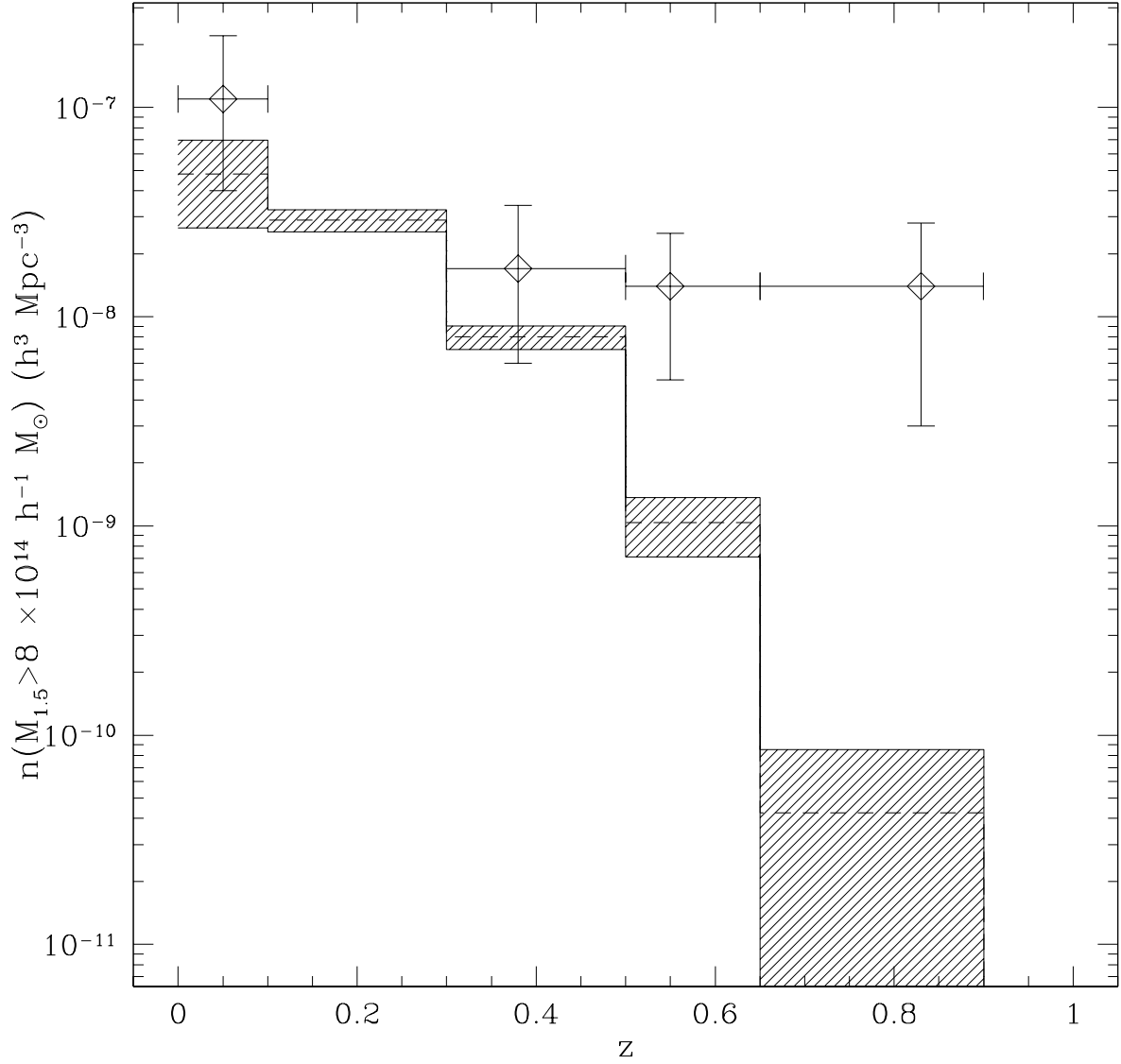}
\caption{The cluster abundance evolution for massive clusters ($M_{1.5} > 
8 \times 10^{14}$ $h^{-1}$ M$_{\odot}$)  in the $\Lambda$CDM model (dashed 
line) versus observed abundances (1-$\sigma$ Poisson error bars).  
}
\end{figure}

\begin{figure}
\label{fig4}
\plotone{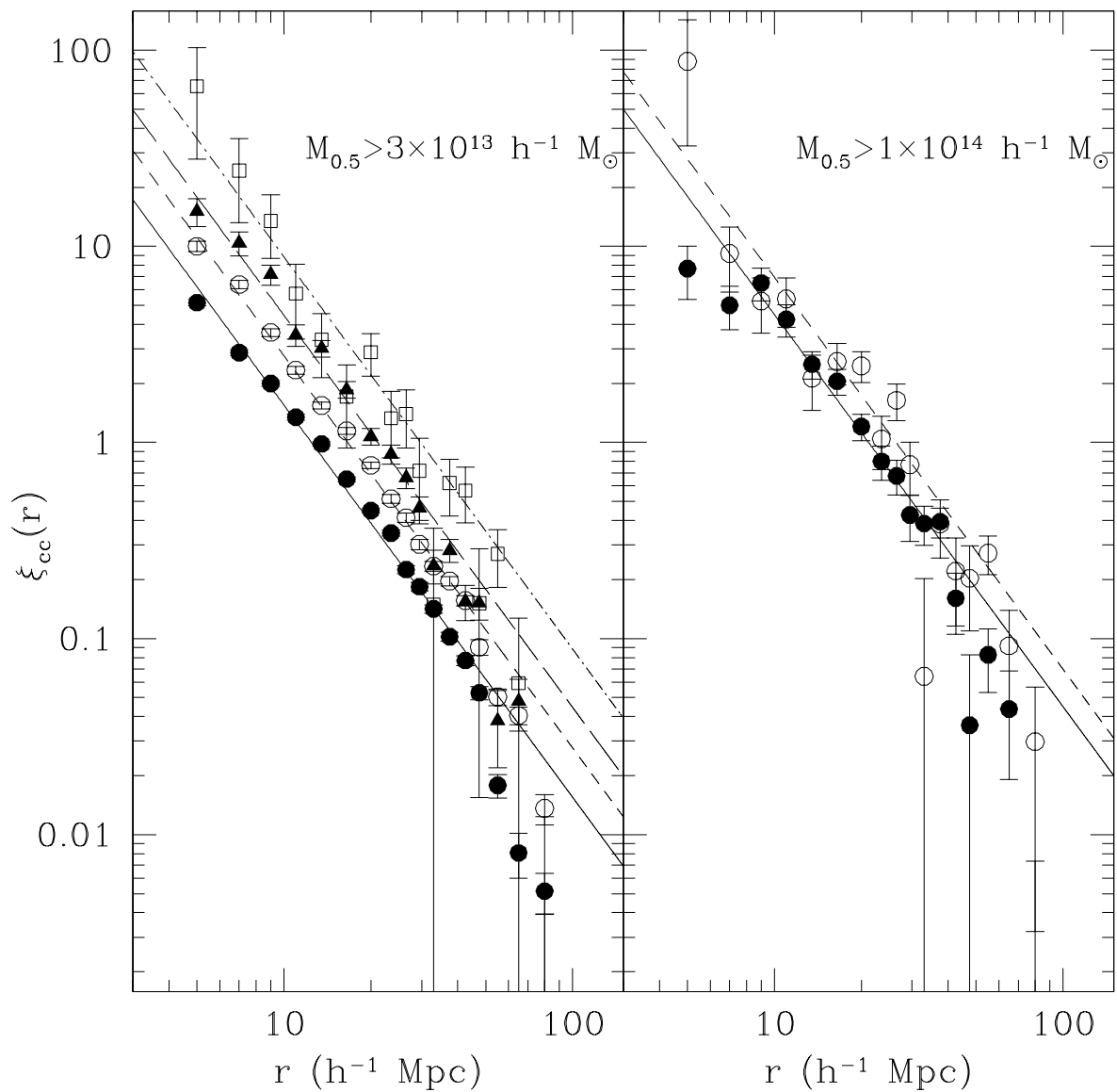}
\caption{The simulated cluster correlation function for different redshift 
bins 
From bottom to top: 
filled circles, $z$=0.0--0.2;
open circles, 0.9--1.1;
filled triangles, 1.4--1.6;
open squares, 1.9--2.2; with 
1-$\sigma$ Poisson error bars.  The separation (r) is in comoving units, and 
the masses are measured within a fixed comoving radius of $0.5$ 
$h^{-1}$ Mpc.  The lines represent the best fit power laws
for a fixed correlation slope $\gamma = 2$.
}
\end{figure}

\begin{figure}
\label{fig5}
\plotone{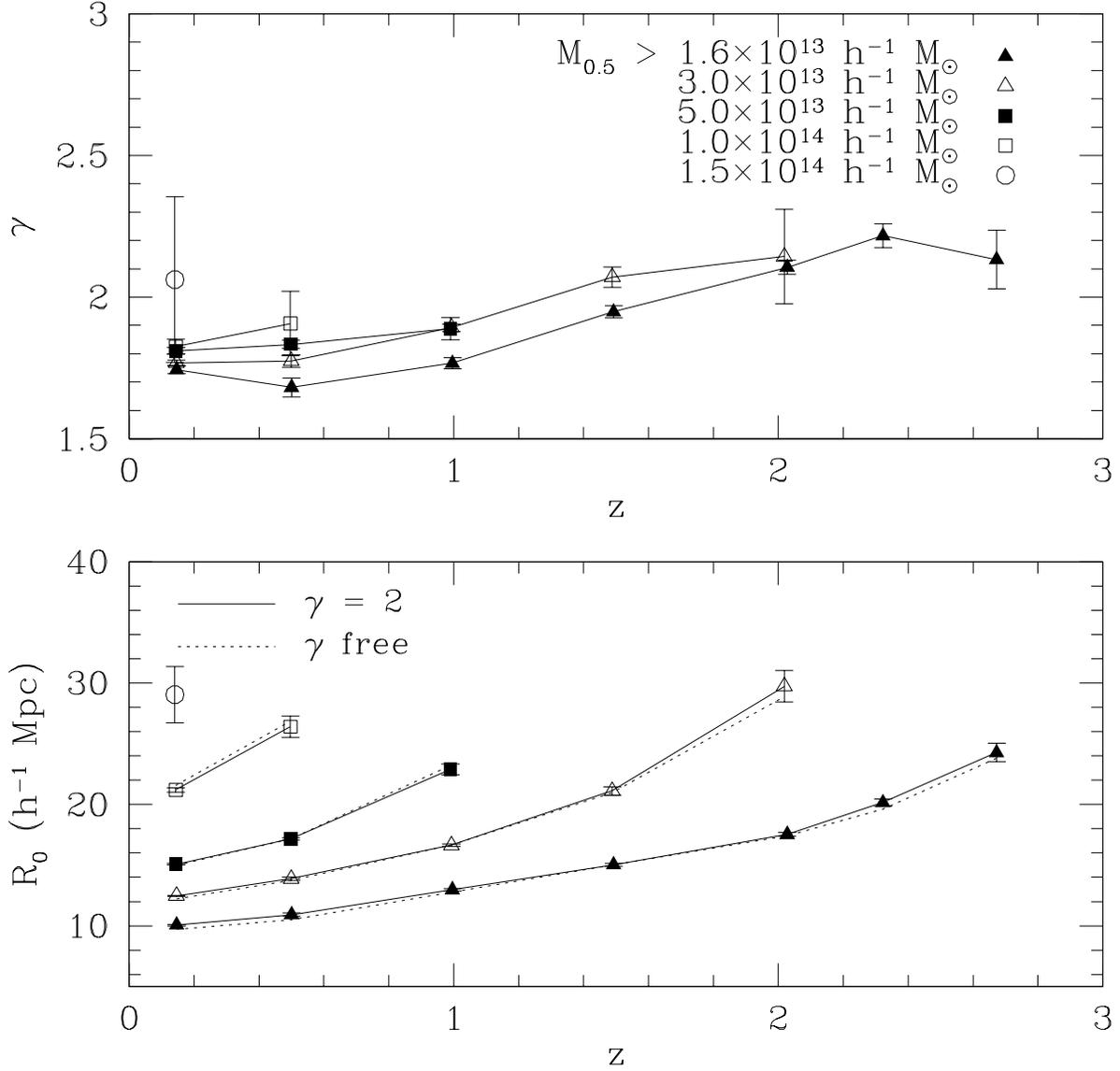}
\caption{Evolution of the cluster correlation scale ($R_0$ comoving) and 
slope ($\gamma$) as a function of redshift for different mass 
thresholds for 
the $\Lambda$CDM model.}
\end{figure}

\begin{figure}
\label{fig6}
\plotone{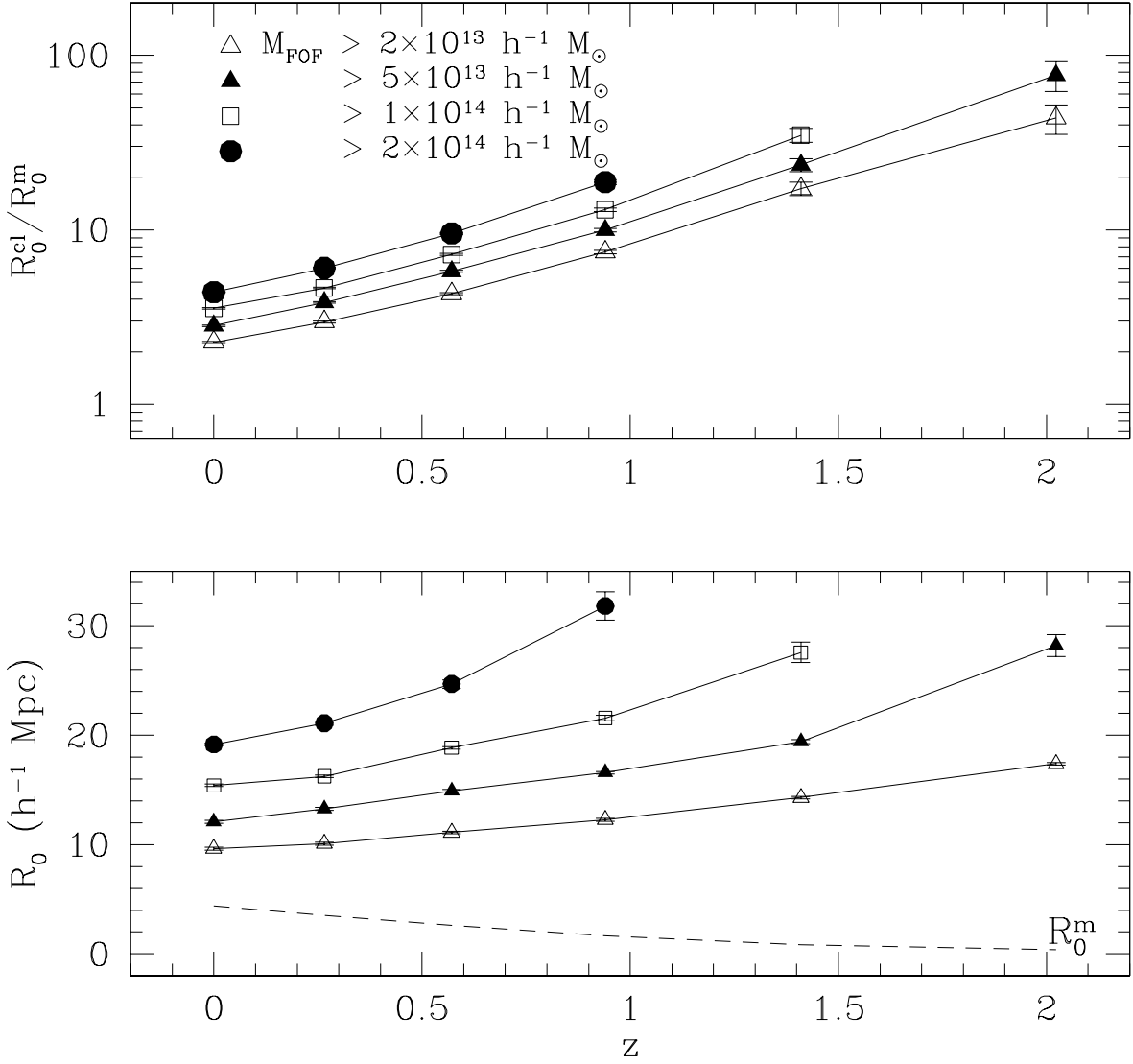}
\caption{The ratio of the correlation scale of clusters ($R_0^{cl}$) to that 
of the underlying mass distribution ($R_0^{m}$) as a function of redshift for 
different mass thresholds.  Also shown are the evolution of $R_0^m$ (dashed 
line) and $R_0^{cl}$, separately.}
\end{figure}

\begin{figure}
\label{fig7}
\plotone{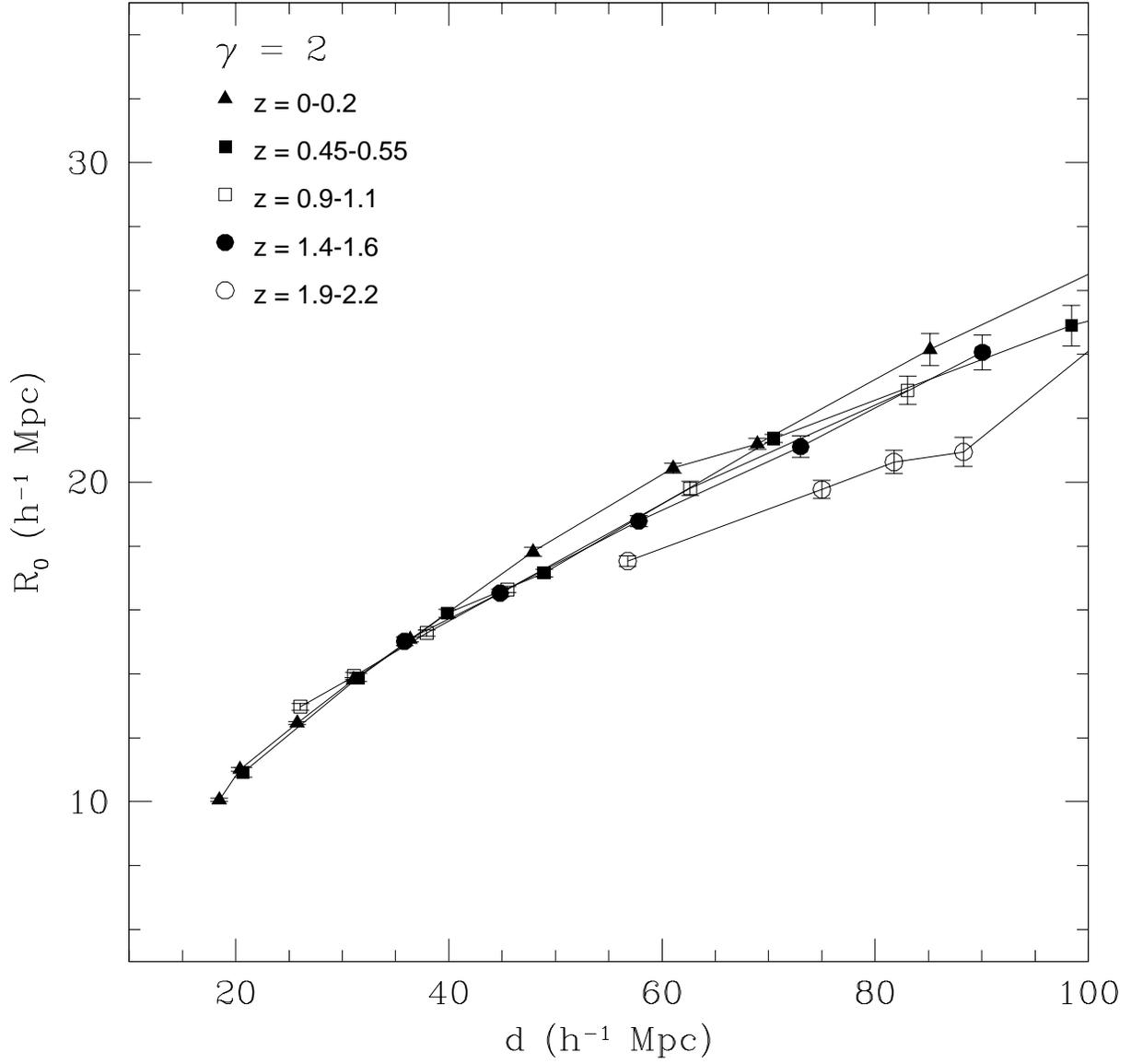}
\caption{The evolution of the richness-dependent cluster correlation function 
: the $R_0-d$ relation (comoving coordinates) for different redshifts using a 
fixed correlation slope $\gamma=2$.}
\end{figure}

\begin{figure}
\label{fig8}
\plotone{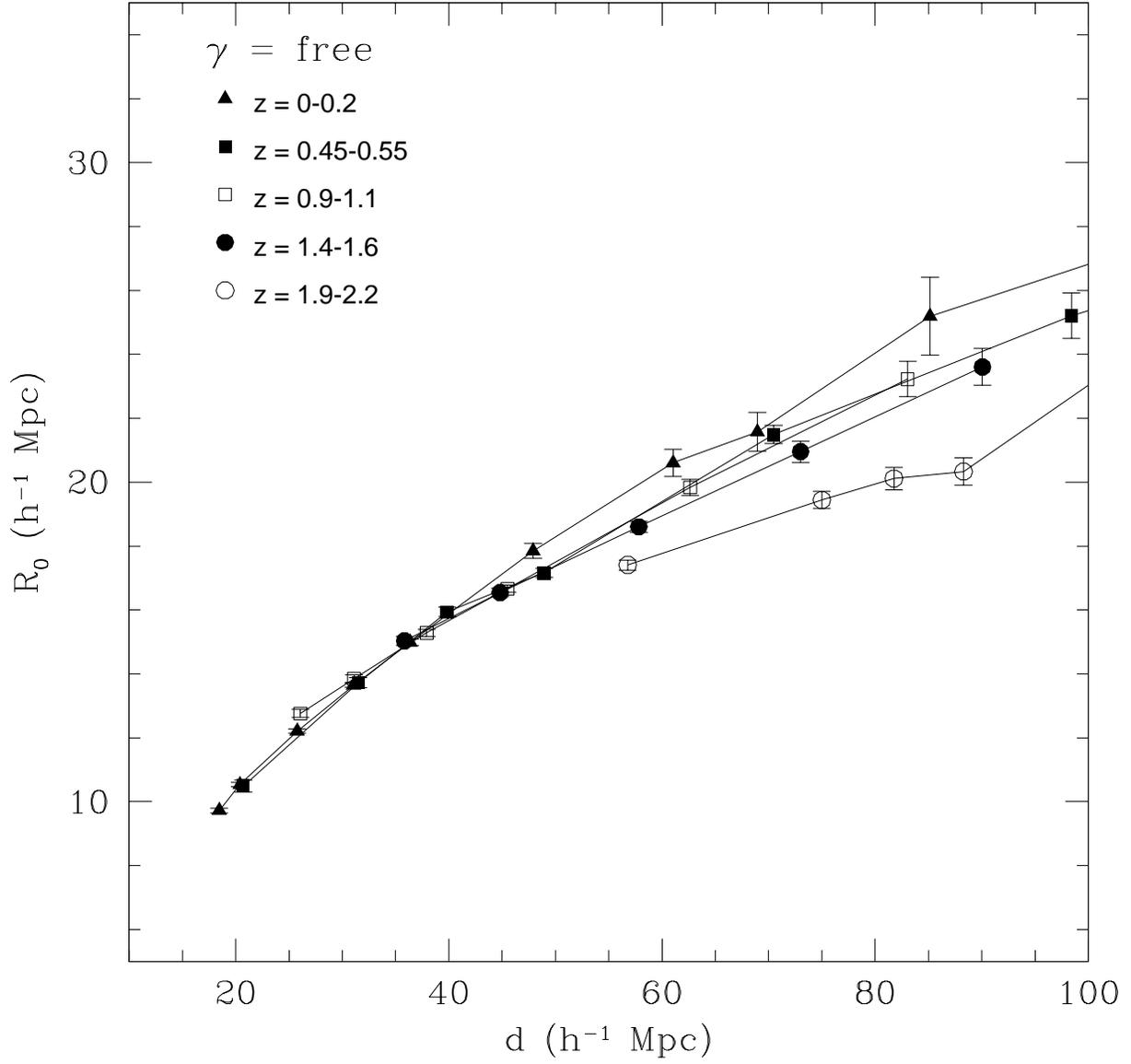}
\caption{The $R_0-d$ relation (comoving coordinates) for different redshifts using a 
free-fitting correlation slope ($\gamma$).}
\end{figure}

\begin{figure}
\label{fig9}
\plotone{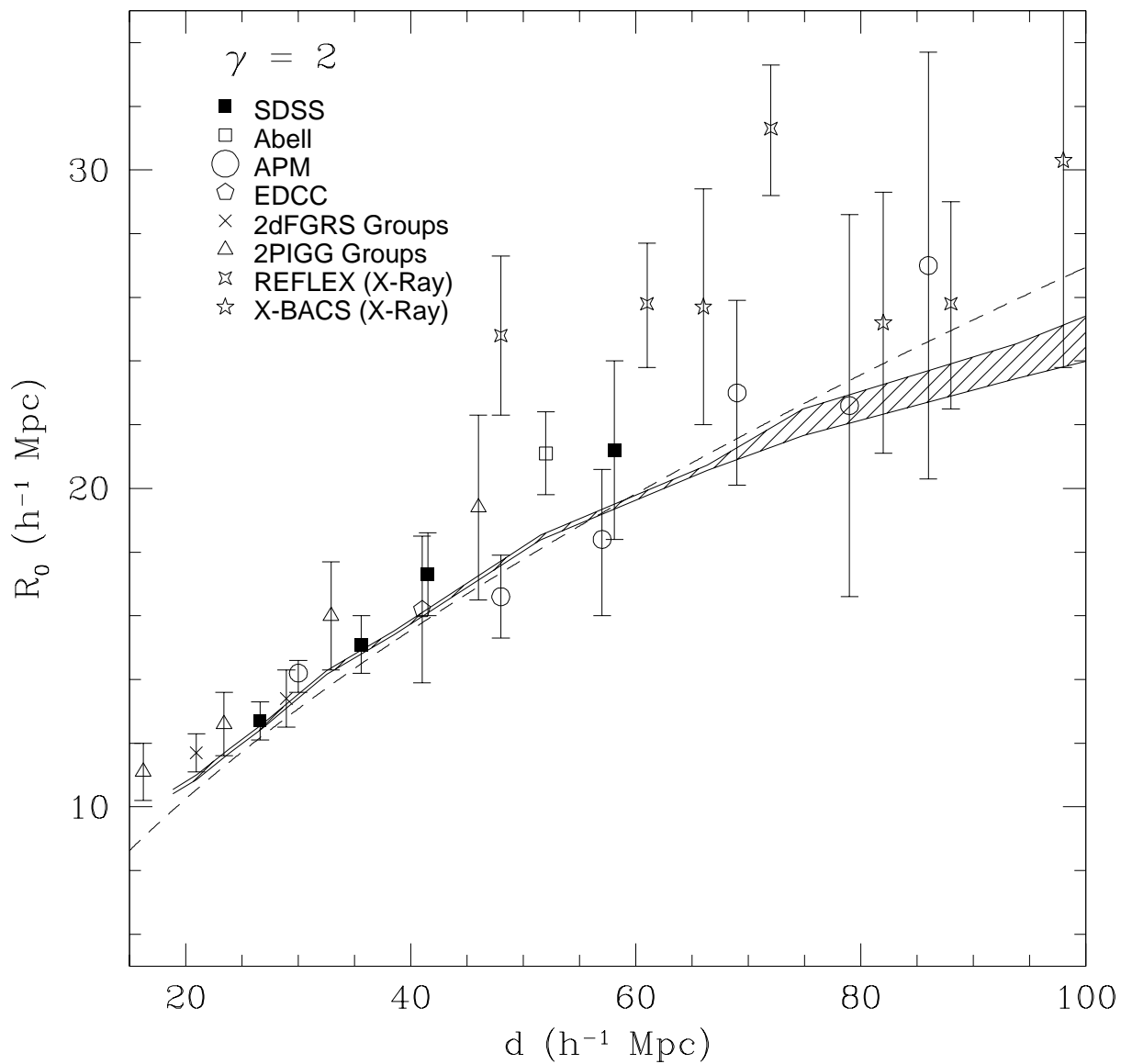}
\caption{A comparison between the $\Lambda$CDM model $R_0-d$ relation
(shaded band; 
$z$=0--0.3) and observational results for a fixed correlation slope $\gamma=2$ 
(data compiled by Bahcall \textit{et al.} 2003c, with the addition 
of the 2PIGG groups from Padilla \textit{et al.} 2004). 
The dashed line is a best fit power law approximation: $R_0 
= 1.7 d^{0.6} h^{-1}$ Mpc.}
\end{figure}

\end{document}